\def\half{\frac{1}{2}}
\def\({\left (}
\def\){\right)}
\def\[{\left [}
\def\]{\right]}
\def\<{\left <}
\def\>{\right>}
\documentclass[a4paper,12pt]{article}
\usepackage{amssymb, amsmath}
\usepackage{epsfig}
\makeatletter
\renewcommand{\section}{{\setcounter{equation}{0}}\@startsection%
{section}%
{1}%
{0mm}%
{-\baselineskip}%
{0.5\baselineskip}%
{\normalfont\normalsize\bfseries}%
} \makeatother
\makeatletter
\renewcommand{\subsection}{\@startsection%
{subsection}%
{2}%
{0mm}%
{-\baselineskip}%
{0.5\baselineskip}%
{\normalfont\normalsize\bfseries}}%
\renewcommand{\theequation}{\arabic{section}.\arabic{equation}}

\newcommand{\ben}{\begin{enumerate}}
\newcommand{\een}{\end{enumerate}}
\newcommand{\be}{\begin{equation}}
\newcommand{\ee}{\end{equation}}
\newcommand{\bea}{\begin{eqnarray}}
\newcommand{\eea}{\end{eqnarray}}
\newcommand{\beas}{\begin{eqnarray*}}
\newcommand{\eeas}{\end{eqnarray*}}
\newcommand{\begth}{\begin{theorem}}
\newcommand{\enth}{\end{theorem}}
\newcommand{\blem}{\begin{lemma}}
\newcommand{\elem}{\end{lemma}}
\newcommand{\non}{\nonumber}
\newcommand{\nl}{\newline}
\newtheorem{remark}{Remark}[section]

\newtheorem{theorem}{Theorem}[section]
\newtheorem{lemma}{Lemma}[section]

\def\RR{\mathbb{R}}

\def\CC{\mathbb{C}}

\def\ZZ{\mathbb{Z}}

\def\la{\langle}
\def\ra{\rangle}
\def\epsq{\epsilon(q)}
\def\sLambda{{\hbox {\tiny{$\Lambda$}}}}
\def\sone{{\hbox {\tiny{$1$}}}}
\def\stwo{{\hbox {\tiny{$2$}}}}

\usepackage{color}

\newcommand{\rd}{\textcolor{red}}

\newcommand{\tit}{\textit}

\begin{document}
\markboth{Models with Recoil for Bose-Einstein Condensation and
Superradiance} {Models with Recoil for Bose-Einstein Condensation
and Superradiance}

\phantom{.} \rd{\textbf{April 24-2005}} \vskip2cm
\begin{center}
{\bf Models with Recoil for Bose-Einstein Condensation and
Superradiance} \vskip 0.5cm {\bf Joseph V.Pul\'e} \footnote{{\tit
Research Associate, School of Theoretical Physics, Dublin
Institute for Advanced Studies.}} \linebreak Department of
Mathematical Physics \linebreak University College
Dublin\\Belfield, Dublin 4, Ireland \linebreak E-mail:
Joe.Pule@ucd.ie

\vskip 0.3cm

{\bf Andr\'{e} F.Verbeure} \linebreak Instituut voor Theoretische
Fysika, \linebreak Katholieke Universiteit Leuven, Celestijnenlaan
200D,\\ 3001 Leuven, Belgium \linebreak E-mail:
andre.verbeure@fys.kuleuven.be

\vskip 0.1cm

and

\vskip 0.1cm

{\bf Valentin A.Zagrebnov} \linebreak Universit\'e de la
M\'editerran\'ee and Centre de Physique Th\'eorique \linebreak
Luminy-Case 907, 13288 Marseille, Cedex 09, France \linebreak
E-mail: zagrebnov@cpt.univ-mrs.fr
\end{center}

\vskip 0.5cm

\begin{abstract}
\vskip -0.7truecm \mbox{}

\noindent In this paper we consider two models which exhibit
\tit{equilibrium} BEC superradiance. They are related to two
different types of superradiant scattering observed in recent
experiments. The first one corresponds to the amplification of
matter-waves due to Raman superradiant scattering from a BE
condensate, when the recoiled and the condensed atoms are in
different internal states. The main mechanism is stimulated Raman
scattering in two-level atoms, which occurs in a superradiant way.
Our second model is related to the superradiant Rayleigh scattering
from a BE condensate. This again leads to a matter-waves
amplification but now with the recoiled atoms in the same state as
the atoms in the condensate. Here the recoiling atoms are able to
interfere with the condensate at rest to form a matter-wave grating
(interference \tit{fringes}) which is observed experimentally.
{\bf  AMS :} 82B10 , 82B23,  81V70

\end{abstract}
\newpage\setcounter{page}{1}

\section{Introduction}
This paper is the third in a series about models for equilibrium
Bose-Einstein Condensation (BEC) superradiance motivated by the
discovery of the Dicke superradiance and BEC matter waves
amplification \cite{S-K}-\cite{KI-2}. In these experiments the
condensate is illuminated with a laser beam, the so called
\tit{dressing beam}. The BEC atoms then scatter photons from this
beam and receive the corresponding recoil momentum producing
coherent \tit{four-wave mixing} of light and atoms \cite{KI-2}.
The aim of our project is the construction of soluble statistical
mechanical models for these phenomena.
\par
In the first paper \cite{PVZ1}, motivated by the principle of
\tit{four-wave mixing} of light and atoms \cite{KI-2}, we
considered two models with a linear interaction between Bose
atoms and photons, one with a global gauge symmetry and another
one in which this symmetry is broken. In both cases we provided a
rigorous proof for the emergence of a cooperative effect between BEC
and superradiance. We proved that there is equilibrium superradiance
and also that there is an enhancement of condensation compared with
that occurring in the case of the free Bose gas.
\par
In the second paper \cite{PVZ2} we formalized the ideas described in
\cite{KI-1,KI-2} by constructing a thermodynamically stable model
whose main ingredient is the \tit{two-level} internal states of
the Bose condensate atoms. We showed that our model is equivalent to
a \tit{bosonized} Dicke maser model. Besides determining its
equilibrium states, we computed and analyzed the thermodynamic
functions, again finding the existence of a cooperative effect
between BEC and superradiance. Here the phase diagram turns out to
be more complex due to the two-level atomic structure.
\par
In the present paper we study the effect of
\tit{momentum recoil} which was omitted in \cite{PVZ1} and
\cite{PVZ2}. Here we consider two models motivated by two
different types of superradiant scattering observed in recent
experiments carried out by the MIT group, see e.g. \cite{S-K}-\cite{K}. Our
first model (\tit{Model 1}) corresponds to the Raman superradiant
scattering from a cigar-shaped BE condensate considered in \cite{S-K}. This
leads to the amplification of matter waves (recoiled atoms) in the
situation when amplified and condensate atoms are in \tit{different}
internal states. The main mechanism is stimulated Raman scattering
in two-level atoms, which occurs in a way similar to Dicke
superradiance \cite{PVZ2}.
\par
Our second model (\tit{Model 2}) is related to the superradiant
Rayleigh scattering from a cigar-shaped BE condensate \cite{I-K},
\cite{K}. This again leads to a matter-wave amplification but now
with recoiled atoms in the \tit{same} state as the condensate
at rest. This is because the condensate is now illuminated by an
off-resonant pump laser beam, so that for a long-pulse the atoms
remain in their lower level states. In this case the
(\tit{non-Dicke}) superradiance is due to self-stimulated Bragg
scattering \cite{K}.
\par
From a theoretical point of view both models are interesting as they
describe homogeneous systems in which there is \tit{spontaneous
breaking of translation invariance}. In the case of the Rayleigh
superradiance this means that the phase transition corresponding to
BEC is at the same time also a transition into a \tit{matter-wave
grating} i.e.
a \lq\lq frozen" spatial density wave structure, see Section 4.
The fact that \tit{recoiling} atoms are able to interfere with
the condensate \tit{at rest} to form a matter-wave grating
(interference \tit{fringes}) has been recently observed
experimentally, see \cite{K}-\cite{KI-2}, and discussion in
\cite{Piov-01} and \cite{Bonif-04}.
\par
In the case of the Raman superradiance there is an important
difference: the internal atomic states for condensed and recoiled
bosons are \tit{orthogonal}. Therefore these bosons are
\tit{different} and consequently cannot interfere to produce a
matter-wave grating as in the first case. Thus the observed spatial
modulation is not in the atomic density of interfering recoiled and
condensed bosons, but in the \tit{off-diagonal} coherence and
photon condensate producing a one-dimensional (\tit{corrugated})
optical lattice, see discussion in Section 4.
\par
Now let us make the definition of our models more exact.
Consider a system of identical bosons of mass $m$ enclosed in a cube
$\Lambda\subset \RR^\nu$ of volume $V = \left|\Lambda\right|$
centered at the origin. We impose periodic boundary conditions so
that the momentum dual set is $\Lambda^*=\{2\pi p /V^{1/\nu}|
p\in\ZZ^\nu\}$.
\par
In \tit{Model 1} the bosons have an internal structure which we
model by considering them as two-level atoms, the two levels being
denoted by $\sigma=\pm$. For momentum $k$ and level $\sigma$,
$a^*_{k,\sigma}$ and $a_{k,\sigma}$ are the usual boson creation and
annihilation operators with
$[a_{k,\sigma},a^*_{k',\sigma'}]=\delta_{k,k'}\delta_{\sigma,\sigma'}$.
Let $\epsilon(k)=\|k\|^2/2m$ be the single particle kinetic energy
and $N_{k,\sigma}=a^*_{k,\sigma}a_{k,\sigma} $ the operator for the
number of particles with momentum $k$ and level $\sigma$. Then the
total kinetic energy is
\begin{equation}\label{kinet1}
T_{\sone,\sLambda}=\sum_{k\in \Lambda^*}\epsilon(k)(N_{k,+}+N_{k,-})
\end{equation}
and the total number operator is $N_{\sone,\sLambda}=\sum_{k\in
\Lambda^*}(N_{k,+}+N_{k,-})$. We define the Hamiltonian
$H_{\sone,\sLambda}$ for \tit{Model 1} by \be
H_{\sone,\sLambda}=T_{\sone,\sLambda}+U_{\sone,\sLambda}
\label{Ham1} \ee where \be U_{\sone,\sLambda}=\Omega\, b^*_q b_q
+\frac{g}{2\sqrt {V}} (a^*_{q+} a_{0 -}b_q +a_{q+} a^*_{0 -}b^*_q
)+\frac{ \lambda}{2{V}} N_{\sone,\sLambda}^2 , \label{Int1} \ee
$g>0$ and $\lambda>0$. Here $b_q$, $b^*_q $ are the creation and
annihilation operators of the photons, which we take as a one-mode
boson field with $[b_q,b^*_q ]=1$ and a frequency $\Omega$. $g$ is
the coupling constant of the interaction of the bosons with the
photon external field which, without loss of generality, we can take
to be positive as we can always incorporate the sign of $g$ into $b$.
Finally the $\lambda$-term is added in
(\ref{Ham1}) to obtain a thermodynamical stable system and to ensure
the right thermodynamic behaviour. This is explained in Section 2.
\par
In \tit{Model 2} we consider the situation when the excited atoms
have already irradiated photons, i.e. we deal only with  de-excited
atoms $\sigma = - $. In other words, we neglect the atom excitation
and consider only \tit{elastic} atom-photon scattering. This is
close to the experimental situation \cite{K}-\cite{KI-2}, in which the
atoms in the BE condensate are irradiated by \tit{off-resonance} laser beam.
Assuming that detuning between the optical fields and the atomic
two-level resonance is much larger that the natural line of the
atomic transition (superradiant Rayleigh regime \cite{I-K, K}) we
get that the atoms always remain in their \tit{lower} internal
energy state. We can then ignore the internal structure of the atoms
and let $a^*_k$ and $a_k$ be the usual boson creation and
annihilation operators for momentum $k$ with
$[a_k,a^*_{k'}]=\delta_{k,k'}$, $N_k=a^*_k a_k $ the operator for
the number of particles with momentum $k$,
\begin{equation}\label{kinet2}
 T_{\stwo,\sLambda}=\sum_{k\in \Lambda^*}\epsilon(k)N_k
\end{equation}
the total kinetic energy, and
$N_{\stwo,\sLambda}=\sum_{k\in\Lambda^*}N_k$ the total number
operator. We then define the Hamiltonian $H_\Lambda^{(2)}$ for
\tit{Model 2} by \be
H_{\stwo,\sLambda}=T_{\stwo,\sLambda}+U_{\stwo,\sLambda}
\label{Ham2} \ee where \be U_{\stwo,\sLambda}=\Omega\, b^*_q b_q
+\frac{g}{2\sqrt {V}} (a^*_{q} a_{0 }b_q +a_{q} a^*_{0}b^*_q
)+\frac{ \lambda}{2{V}}N_{\stwo,\sLambda}^2\,. \label{Int-2} \ee
\par
This paper is structured as follows: \\
In Section 2 we give a complete rigorous solution of the variational principle
for the equilibrium state for \tit{Model 1} and we compute the corresponding
pressure as a function of the temperature and the chemical
potential. We prove that this model exhibits
Raman superradiance. \\
In Section 3 we study \tit{Model 2} and show that Rayleigh superradiance occurs in this case.
The analysis is very similar to that of \tit{Model 1} and therefore we do not repeat it but simply state
the results. \\
In Section 4  we show that in both models there is spontaneous breaking of translation
invariance in the equilibrium state. We relate this with the spatial modulation of
matter-waves (\tit{matter-wave grating}). We find that in \tit{Model 1}
there is no such spatial modulation
in spite of the breaking of translation invariance while in \tit{Model 2}
this spatial modulation exists.
We conclude with several remarks.
\par
We close this introduction with the following comments:\\
- In our models we \tit{do not} use for effective photon-boson
interaction the \tit{four-wave mixing principle}, see
\cite{KI-2}, \cite{PVZ1}, \cite{M-M1}. The latter seems to be
important for the geometry, when a linearly polarized pump laser
beam is incident in a direction perpendicular to the long axis of
a cigar-shaped BE condensate, inducing the $\lq\lq 45^{\circ}$-
recoil pattern" picture \cite{S-K}-\cite{K}. Instead as in
\cite{PVZ2}, we consider a \tit{minimal} photon-atom
interaction only with superradiated photons, cf \cite{M-M2}. This
corresponds to superradiance in a \lq\lq one-dimensional"
geometry, when a pump laser beam is collimated and aligned along
the long axis of
a cigar-shaped BE condensate, see \cite{Bonif-04}, \cite{Koz}.\\
- In this geometry the superradiant photons and recoiled
matter-waves propagate in the same direction as the incident pump
laser beam. If one considers it as a classical \lq\lq source" (see
\cite{KI-2}), then we get a  \tit{minimal} photon-atom
interaction \cite{PVZ2} generalized to take into account the effects
of recoil. Notice that the further approximation of the BEC
operators by c-numbers leads to a
bilinear photon-atom interaction studied in \cite{KI-2}, \cite{PVZ1}.\\
- In this paper we study \tit{equilibrium} BEC superradiance
while the experimental situation (as is the case with Dicke
superradiance \cite{D}) is more  accurately described by
non-equilibrium statistical mechanics. However we believe that for
the purpose of understanding the quantum coherence interaction
between light and the BE condensate our analysis is as instructive
and is in the same spirit as the rigorous study of the Dicke model
in thermodynamic equilibrium, see e.g.
\cite{HL}-\cite{FSVW}.\\
- In spite of the simplicity of our exactly soluble \tit{Models
1} and \tit{2} they are able to demonstrate the main features
of the BEC superradiance with recoil: the photon-boson condensate
\tit{enhancement} with formation of the \tit{light
corrugated optical lattice} and the \tit{matter-wave grating}.
The corresponding phase diagrams are very similar to those in
\cite{PVZ2}. However, though the type of behaviour is similar,
this is now partially due to the momentum recoil and not entirely
to the internal atomic level structure.
\section{Model 1}
\subsection{The effective Hamiltonian}
\setcounter{equation}{0}
\renewcommand{\theequation}{\arabic{section}.\arabic{equation}}
\setcounter{theorem}{0}

We start with the \tit{stability} of  Hamiltonian (\ref{Ham1}).
Consider the term $U_{\sone,\sLambda}$ in (\ref{Int1}). This gives
\bea U_{\sone,\sLambda} &=&\Omega\, ( b^*_q +\frac{g}{2\Omega\sqrt
{V}} a_{q +} a^*_{0 -})(b_q + \frac{g}{2\Omega\sqrt {V}} a^*_{q  +}
a_{0 -})
-\frac{g^2}{4 \Omega^2 {V}} N_{0-}(N_{q  +}+1)+\frac{ \lambda}{2{V}} N_{\sone,\sLambda}^2\non\\
&\geq & \frac{ \lambda}{2{V}} N_{\sone,\sLambda}^2-\frac{g^2}{4 \Omega^2
{V}} N_{0-}(N_{q  +}+1). \label{lbound} \eea On the basis of the
trivial inequality $4ab\leq (a+b)^2$, the last term in the lower
bound in (\ref{lbound}) is dominated by the first term if $\lambda
> g^2/8\Omega $, that is if the stabilizing coupling $\lambda$
is large with respect to the coupling constant $g$ or if the
external frequency is large enough. We therefore assume the
\tit{stability condition}: $\lambda
> g^2/8\Omega $.
\par
Since we want to study the equilibrium
properties of the model (\ref{Ham1}) in the grand-canonical
ensemble, we shall work with the Hamiltonian
\be
H_{\sone,\sLambda}(\mu)=H_{\sone,\sLambda}-\mu N_{\sone,\sLambda}
\label{Ham-mu}
\ee where
$\mu$ is the chemical potential. Since $T_{\sone,\sLambda}$ and the
interaction $U_{\sone,\sLambda}$ conserve the quasi-momentum, Hamiltonian
(\ref{Ham1}) describes a \tit{homogeneous} (translation
invariant) system. To see this explicitly, notice that the
external laser field possesses a natural quasi-local structure as
the Fourier transform of the field operator $b(x)$:
\begin{equation}\label{bq}
b_q=\frac{1}{\sqrt V}\int_\Lambda dx\, e^{iq\cdot x}b(x).
\end{equation}
If for $z\in\RR^\nu$, we let $\tau_x$ be the translation
automorphism $(\tau_z b)(x)= b(x + z)$, then since we have periodic
boundary conditions, $\tau_z
(b_q)=e^{-iq\cdot z}b_q$ and similarly $\tau_z
(a_{k,\sigma})=e^{-ik\cdot z}a_{k,\sigma}$. Therefore, the
Hamiltonian (\ref{Ham1}) is translation invariant.
Consequently, in the thermodynamic limit, it is natural to
look  for translation invariant or homogeneous equilibrium states at
all inverse temperatures $\beta$ and all values of the chemical
potential $\mu$.
\par
Because the interaction (\ref{Int1}) is not bilinear or quadratic in
the creation and annihilation operators the system cannot be
diagonalized by a standard symplectic or Bogoliubov transformation.
Therefore at the first glance one is led to conclude that the model
is not soluble. However on closer inspection one notices that all
the interaction terms contain  space averages, namely, either
\begin{equation}\label{a-0}
\frac{a_{0-}}{\sqrt V \phantom{..}}=\frac{1}{V}\int_\Lambda
dx\,a_-(x),
\end{equation}
and its adjoint, or \be \frac{1}{V}\int_\Lambda dx\, a^*_\sigma(x)
a_\sigma(x). \ee Without going into all the mathematical details it
is well-known \cite{BR} that space averages tend weakly to a
multiple of the identity operator for all space-homogeneous
\tit{extremal} or \tit{mixing} states. Moreover as all the
methods of characterizing the equilibrium states (e.g. the
variational principle, the KMS-condition, the characterization by
correlation inequalities etc. \cite{BR}) involve only affine
functionals on the states, we can limit ourselves to looking for the
extremal or mixing equilibrium states and in so doing we can exploit
the above mentioned property for space averages. One way of
accomplishing this is by applying the so-called \tit{effective
Hamiltonian} method, which is based on the fact that an equilibrium
state is not determined by the Hamiltonian but by its
\tit{Liouvillian}. The best route to prove the exactness
of the effective Hamiltonian method (cf
\cite{FSV}) is to use the characterization of the equilibrium state
by means of the \tit{correlation inequalities}
\cite{FV}, \cite{BR}:\\
{\tit A state $\omega$ is an equilibrium state for
$H_{\sone,\sLambda}(\mu)$ at inverse temperature $\beta$, if and
only if for all local observables $A$, it satisfies} \be \lim_{V \to
\infty}\beta\omega\([A^*,[H_{\sone,\sLambda}(\mu),
A]]\)\geq\omega(A^*A)\ln \frac{\omega(A^*A)}{\omega(AA^*)}.
\label{corr ineq} \ee Clearly only the \tit{Liouvillian}
$[H_{\sone,\sLambda}(\mu), \cdot]$ of the Hamiltonian enters into
these inequalities and therefore we can replace
$H_{\sone,\sLambda}(\mu)$ by a simpler Hamiltonian, the {\tit
effective Hamiltonian}, which gives in the limiting state $\omega$
the same Liouvillian as $H_{\sone,\sLambda}(\mu)$ and then look for
the equilibrium states corresponding to it. Now in our case for an
extremal or mixing state $\omega$ we define the effective
translation invariant Hamiltonian $H_{\sone,\sLambda}^{\rm
eff}(\mu,\eta,\rho)$ such that for all local observables $A$ and $B$
\be \lim_{V \to \infty}\omega\(A,[H_{\sone,\sLambda}(\mu),
B]\)=\lim_{V \to \infty}\omega\(A,[H_{\sone,\sLambda}^{\rm
eff}(\mu,\eta,\rho), B]\). \label{eff comm} \ee The significance of
the \tit{parameters} $\eta$ and $\rho$ will become clear below.
One can then replace (\ref{corr ineq}) by \be \lim_{V \to
\infty}\beta\omega\([A^*,[H_{\sone,\sLambda}^{\rm
eff}(\mu,\eta,\rho), A]]\)\geq\omega(A^*A)\ln
\frac{\omega(A^*A)}{\omega(AA^*)}. \label{corr ineq eff} \ee We can
choose $H_{\sone,\sLambda}^{\rm eff}(\mu,\eta,\rho)$ so that it can
be diagonalized and thus (\ref{corr ineq eff}) can be solved
explicitly. For a given chemical potential $\mu$, the inequalities
(\ref{corr ineq eff}) can have \tit{more than one} solution. We
determine the physical solution by minimizing the free energy
density with respect to the set of states or equivalently by
maximizing the grand canonical pressure on this set.\\
Let the effective Hamiltonian  be defined by
\bea
\label{eff-Ham-0} H_{\sone,\sLambda}^{\rm eff}(\mu,\eta,\rho)
&=&(\lambda \rho-\mu+\epsq) a^*_{q+}a_{q+}+(\lambda \rho-\mu)
a^*_{0-}a_{0-}+\frac{g}{2}(\eta  a^*_{q+}b_q +{\bar \eta }a_{q+}b^*_q)\non\\
&&\hskip 1.5cm +\Omega\, b^*_qb_q+\frac{g\sqrt V}{2}\ ({\bar \zeta}
a_{0-}+ \zeta a^*_{0-}) + T'_{\sone,\sLambda} +(\lambda \rho-\mu)
N'_{\sone,\sLambda} \eea where \be T'_{\sone,\sLambda}=\sum_{k\in
\Lambda^*,\,k\neq q}\epsilon(k)N_{k,+}+\sum_{k\in \Lambda^*,\,k\neq
0}\epsilon(k)N_{k,-}, \ee \be N'_{\sone,\sLambda}=\sum_{k\in
\Lambda^*,\,k\neq q}N_{k,+}+\sum_{k\in \Lambda^*,\,k\neq 0}N_{k,-},
\ee $\eta$ and $\zeta$ are complex numbers and $\rho$ is a positive
real number. Notice that the Hamiltonian (\ref{eff-Ham-0}) is
translation invariant, but it is not gauge invariant for $\zeta \neq
0$ because of the linear terms in $a^*_{0-}, a_{0-}$ operators.
Therefore in this case, (\ref{eff-Ham-0}) generates translation
invariant states, which are extremal with respect to the gauge group
in the zero-minus mode. They are labeled by the $\arg\,\zeta $. One
can easily check that (\ref{eff comm}) is satisfied if \be
\eta=\frac{\omega(a_{0-} )}{\sqrt V},\ \ \ \ \zeta
=\frac{\omega(a_{q+} b_{q}^{*})}{V}  \ \ \ \ {\rm and}\ \ \ \ \rho
=\frac{\omega(N_{\sone,\sLambda})}{V}, \label{consist} \ee where the
state $\omega $ coincides with the equilibrium state $\la\, \cdot\,
\ra_{H_{\sone,\sLambda}^{\rm eff}(\mu,\eta,\rho)}$ defined by the
effective Hamiltonian $H_{\sone,\sLambda}^{\rm eff}(\mu,\eta,\rho)$.
From (\ref{eff comm}) and (\ref{consist}) we then obtain the
\tit{self-consistency} equations \be \eta =\frac{1}{\sqrt V}\la
a_{0-} \ra_{H_{\sone,\sLambda}^{\rm eff}(\mu,\eta,\rho)},\ \ \ \
\zeta =\frac{1}{V}\la a_{q+} b_q ^*\ra_{H_{\sone,\sLambda}^{\rm
eff}(\mu,\eta,\rho)} ,\ \ \ \ \rho =\frac{1}{V}\la
N_{\sone,\sLambda}\ra_{H_{\sone,\sLambda}^{\rm eff}(\mu,\eta,\rho)}.
\label{consist0} \ee The structure of (\ref{eff-Ham-0}) implies that
the parameter $\zeta $ is a function of $\eta $ and $\rho$ through
(\ref{consist0}). So, we do not need to label the effective
Hamiltonian by $\zeta $. The important simplification here is that
$H_{\sone,\sLambda}^{\rm eff}(\mu,\eta,\rho)$ can be diagonalized:
\bea \label{eff-Ham-1} H_{\sone,\sLambda}^{\rm
eff}(\mu,\eta,\rho)&=&E_{+}
(\mu,\eta,\rho)\alpha^*_1\alpha_1+E_{-}(\mu,\eta,\rho)\alpha^*_2\alpha_2  \\
&&\hskip 1cm +(\lambda \rho-\mu) \alpha^*_3\alpha_3
+T'_{\sone,\sLambda}+(\lambda \rho-\mu) N'_{\sone,\sLambda}
+\frac{g^2V |\zeta |^2}{4(\mu -\lambda \rho)}, \non
\eea where
\bea
E_{+}(\mu,\eta,\rho)&=& \half(\Omega-\mu+\lambda\rho +\epsq)+
\half\sqrt{(\Omega+\mu-\lambda\rho -\epsq)^2+g^2|\eta|^2},\non\\
E_{-}(\mu,\eta,\rho)&=& \half(\Omega-\mu+\lambda\rho +\epsq)-
\half\sqrt{(\Omega+\mu-\lambda\rho -\epsq)^2+g^2|\eta|^2},\non\\
\eea \be \alpha_1=a_{q+}\cos \theta  +b_q\sin \theta ,\ \ \ \ \
\alpha_2=a_{q+}\sin \theta  -b_q\cos \theta,\ \ \ \ \ \alpha_3=a_{0
-}+\frac{g\sqrt{V}\zeta }{2(\lambda \rho-\mu)}, \ee and \be \tan
2\theta =-\frac {g|\eta|}{\Omega+\mu -\lambda \rho-\epsq}. \ee Note
that the correlation inequalities (\ref{corr ineq eff}) (see
\cite{FV}) imply that \be \lim_{V \to
\infty}\omega\(A^*,[H_{\sone,\sLambda}(\mu), A]\)\geq 0
\label{positive} \ee for all observables $A$. Applying
(\ref{positive}) with $A=a^*_{0+}$, one gets the condition
$\lambda \rho -\mu\geq 0 $. Similarly, one obtains the condition
$\lambda \rho + \epsq -\mu\geq 0 $ by applying (\ref{positive}) to
$A=a^*_{q+}$. We also have that $E_{+}(\mu,\eta,\rho)\geq
E_{-}(\mu,\eta,\rho)$ and $E_{-}(\mu,\eta,\rho)=0$ when
$|\eta|^2=4\Omega(\lambda \rho+\epsq-\mu)/g^2 $ and then
$E_{+}(\mu,\eta,\rho)=\Omega-\mu +\lambda \rho+\epsq$. Thus we have
the constraint:
\begin{equation}\label{eta-inequal}
|\eta|^2 \leq 4\Omega(\lambda \rho+\epsq-\mu)/g^2
\end{equation}
We shall need the above information to make sense of the
thermodynamic functions below. Of course the parameters $\eta$,
$\rho$ and consequently $E_\pm$ are $V$ dependent but for simplicity
we do not indicate this dependence explicitly.
\par
We can foresee that for some values of $\mu$ there will be
Bose-Einstein condensation in the mode $\{0-\}$. We know that in
this case the gauge invariant, homogeneous states are not extremal
within the class of translation invariant equilibrium states
\cite{FPV}. Therefore to ensure that the states that we shall obtain
are extremal we add to the Hamiltonian a gauge breaking term
\begin{equation}\label{h-sources}
-\frac{g \sqrt V}{2} \(\overline{h}a_{0-}+ h a^*_{0-}\) ,
\end{equation}
and then let $h \rightarrow 0$ after the thermodynamic limit. The
corresponding effective Hamiltonian is then \be \label{eff-Ham-s}
H_{\sone,\sLambda}^{\rm eff}(\mu,\eta,\rho; h):=
H_{\sone,\sLambda}^{\rm eff}(\mu,\eta,\rho)- \frac{g \sqrt V}{2}
\(\overline{h}a_{0-}+ h a^*_{0-}\) \ee with $h\in\CC$. The equations
corresponding to (\ref{consist0}) now become \be \eta
=\frac{1}{\sqrt V}\la a_{0-} \ra_{H_{\sone,\sLambda}^{\rm
eff}(\mu,\eta,\rho;h)},\ \ \ \ \zeta =\frac{1}{V}\la a_{q+} b_q
^*\ra_{H_{\sone,\sLambda}^{\rm eff}(\mu,\eta,\rho;h)} , \ \ \ \ \rho
=\frac{1}{V}\la N_{\sone,\sLambda}\ra_{H_{\sone,\sLambda}^{\rm
eff}(\mu,\eta,\rho;h)}. \label{consist1a} \ee These consistency
equations can be made explicit by using the above diagonalization:
\be \eta=\frac{g}{2(\mu -\lambda \rho)}(\zeta - h), \label{A} \ee
\be \zeta=\half \frac{g\,\eta}{V(E_+-E_-)}\left\{\frac{1}{e^{\beta
E_+}-1}-\frac{1}{e^{\beta E_-}-1} \right\} \label{B} \ee and \bea
\rho &=&|\eta|^2+\frac{1}{V}\frac{1}{e^{-\beta(\mu -\lambda
\rho)}-1}+
\frac{1}{2V}\left\{\frac{1}{e^{\beta E_+}-1}+\frac{1}{e^{\beta E_-}-1} \right\}\non\\
&&-\frac{(\mu -\lambda \rho -\epsq +\Omega)}{2 V(E_+-E_-)}
\left\{\frac{1}{e^{\beta E_+}-1}-\frac{1}{e^{\beta E_-}-1} \right\}\non\\
&& +\frac{1}{V}\sum_{k\in \Lambda^*,\,k\neq \,0} \frac{1}{e^{\beta
(\epsilon (k)-\mu +\lambda \rho)}-1}
 +\frac{1}{V}\sum_{k\in \Lambda^*,\,k\neq \,0,\,q} \frac{1}{e^{\beta (\epsilon (k)-\mu +\lambda \rho)}-1}.
\label{density1}
\eea
Combining (\ref{A}) and (\ref{B}) we obtain
the equation:
\be
\eta =\frac{g^2 \,\eta}{4(\mu
-\lambda \rho) V(E_+-E_-)} \left\{\frac{1}{e^{\beta
E_+}-1}-\frac{1}{e^{\beta E_-}-1} \right\}-\frac{gh}{2(\mu -\lambda \rho)}.
\label{consist1}
\ee
It is now clear that the equilibrium states are determined by the
limiting form of the consistency equations (\ref{A}) - (\ref{consist1}).
We solve these equations and obtain the
corresponding pressure so that we can determine the equilibrium
state, when there are several solutions for a particular chemical
potential.
\par
We shall need the following definitions:
\be
\varepsilon_0(\mu)=\frac{1}{(2\pi)^3}\int_{\RR^3}d^3k\frac{\epsilon
(k)-\mu}{e^{\beta (\epsilon (k)-\mu)}-1},
\ee
\be
\rho_0(\mu)=\frac{1}{(2\pi)^3}\int_{\RR^3}d^3k\frac{1}{e^{\beta
(\epsilon (k)-\mu)}-1}
\ee
and
\be
p_0(\mu)=-\frac{1}{(2\pi)^3}\int_{\RR^3}d^3k\ln(1-e^{-\beta
(\epsilon (k)-\mu)}),
\ee
that is the grand-canonical energy
density, the particle density and the pressure for the \tit{free
Bose-gas} for $\mu \leq 0$. Let
\be
s_0(\mu)=\beta(\varepsilon_0(\mu)+p_0(\mu)),
\ee
and note that $s_0(\mu)$ is an increasing function of $\mu$. We
shall denote the free Bose-gas \tit{critical} density by
$\rho_c$, i.e. $\rho_c: = \rho_0(0)$. Recall that $\rho_c$ is
infinite for $\nu<3$ and finite for $\nu\geq 3$.

\subsection{Solution of consistency equations}

Notice first that equations (\ref{A}) and (\ref{B}) imply
\begin{equation}\label{phase}
\arg \zeta = \arg \eta = \arg h  = \varphi\,,
\end{equation}
i.e. one can consider the corresponding parameters in the
consistency equations (\ref{A}) - (\ref{consist1}) to be real and
non-negative.
\begin{remark}\label{rem-2.2}
By virtue of the upper bound (\ref{eta-inequal}) and equation
(\ref{consist1}) we get that $\delta:=\lambda \rho - \mu
> 0$ for any volume $V$ as soon as $h \neq 0$. By the same reasoning one gets
from (\ref{consist1}) that  $\lim_{V\to\infty} \inf E_{-} > 0$ (Case
A), or at most $\lim_{V\to\infty} E_{-} =0$ in such a way that
$\lim_{V\to\infty} V E_{-}$ is finite (Case B).
\end{remark}
We start our analysis of the solution of the consistency equations
from small densities (small chemical potentials), when there is no
condensates, passing then to higher values. So, later on we
distinguish the two possibilities indicated in Remark \ref{rem-2.2}:
\begin{equation}\label{Case-A}
Case\,\, A: \,\,\,\,\,\,\,\,\,\,\,\,\,\,\,\, \lim_{V\to\infty} E_{-}
> 0 .
\end{equation}
Then by Remark \ref{rem-2.2} the consistency equations (\ref{A}) -
(\ref{consist1}) in the thermodynamic limit yield:
\begin{equation}\label{lim-small-mu}
\eta = \frac{gh}{2\delta}\,\,\,,\,\,\,\, \zeta = 0 \,,
\end{equation}
and the equation for the particle density (\ref{density1}) takes the
form:
\begin{equation}\label{density11}
\mu=2\lambda\rho_0(-\delta)-\delta+\frac{\lambda
g^2|h|^2}{4\delta^2} .
\end{equation}
We also have the limiting expectations \be
\lim_{V\to\infty}\frac{1}{V}\la
a^*_{0-}a_{0-}\ra_{H_{\sone,\sLambda}^{\rm eff}(\mu,\eta,\rho;h)}
=|\eta|^2=\frac{g^2|h|^2}{4\delta^2} \label{cond1} \ee and (by
virtue of (\ref{Case-A})) \be \lim_{V\to\infty}\frac{1}{V}\la
a^*_{q+}a_{q+}\ra_{H_{\sone,\sLambda}^{\rm eff}(\mu,\eta,\rho;h)}
=\lim_{V\to\infty}\frac{1}{V}\la b^*_q
b_q\ra_{H_{\sone,\sLambda}^{\rm eff}(\mu,\eta,\rho;h)} =0.
\label{cond2} \ee Let us now examine equation (\ref{density11}). For
$h \neq 0$ it is  solvable for any $\mu$ and we shall denote its
unique solution by $\delta(\mu,h)= \lambda \rho(\mu,h) - \mu$.

\textbf{Solution 1:}\ \ Suppose that $\mu\leq \mu_c:=2\lambda\rho_c
\equiv 2\lambda\rho_0 (0)$. Since $|h|/\delta\to 0$ as $h\to 0$,
then  $\delta(\mu,h)\to \delta(\mu)$, where $\delta(\mu)$ is the
unique solution of equation: \be \mu=2\lambda\rho_0(-\delta)-\delta
. \label{density12} \ee Then we see from (\ref{cond1}) and
(\ref{cond2}) that in this case in the thermodynamic limit there is
\tit{no condensation} in the $\{0-\}$ and other modes: \be
\lim_{V\to\infty}\frac{1}{V}\la
a^*_{0-}a_{0-}\ra_{H_{\sone,\sLambda}^{\rm eff}(\mu,0,\rho)}
=\lim_{V\to\infty}\frac{1}{V}\la
a^*_{q+}a_{q+}\ra_{H_{\sone,\sLambda}^{\rm eff}(\mu,0,\rho)}
=\lim_{V\to\infty}\frac{1}{V}\la b^*_q
b_q\ra_{H_{\sone,\sLambda}^{\rm eff}(\mu,0,\rho)} =0. \ee After
letting $h\to 0$, the energy density is given by \bea
\lim_{V\to\infty}\frac{1}{V}\la H_{\sone,\sLambda}
(\mu)\ra_{H_{\sone,\sLambda}^{\rm eff}(\mu,0,\rho)} &=&
2\varepsilon_0(-\delta(\mu)) -2(\delta(\mu)+\mu)
\rho_0(-\delta(\mu))
+\half \lambda (\delta(\mu)+\mu)^2\non\\
&=& 2\varepsilon_0(-\delta(\mu))
-\half \frac{(\delta(\mu)+\mu)^2} {\lambda}
\eea
and the entropy density is equal to
\be
s(\mu)=2s_0(-\delta(\mu)).
\ee
Since the grand-canonical
pressure is given by
\be
p(\mu)=\frac{1}{\beta}s(\mu)
-\lim_{V\to\infty}\frac{1}{V}\la H_{\sone,\sLambda}
(\mu)\ra_{H_{\sone,\sLambda}^{\rm eff}(\mu,\eta,\rho)},
\ee then
\be
p(\mu)=2p_0(-\delta(\mu))+\half \frac{(\delta(\mu)+\mu)^2} {\lambda} .
\ee
\textbf{Solution 2:}\ \ Now suppose that $\mu>\mu_c = 2\lambda
\rho_c$. Then to verify equation  (\ref{density11}) in the limit
$h\to 0$ the solution must converge to zero: $\delta(\mu,h)\to 0$,
in such a way that \be \frac{\lambda g^2|h|^2}{4\delta^2(\mu,h)}\to
\mu - 2\lambda \rho_c. \ee Therefore it follows from
(\ref{density11}) and (\ref{cond1}) that
\begin{equation}\label{eta-case-2}
|\eta|^2=\lim_{V\to\infty}\frac{1}{V}\la
a^*_{0-}a_{0-}\ra_{H_{\sone,\sLambda}^{\rm
eff}(\mu,\eta,\rho)}=\frac{\mu}{\lambda} -2\rho_c
\end{equation}
and (again by (\ref{Case-A})) the limit (\ref{cond2}) gives
\be\label{case-2} \lim_{V\to\infty}\frac{1}{V}\la
a^*_{q+}a_{q+}\ra_{H_{\sone,\sLambda}^{\rm eff}(\mu,\eta,\rho)} =
\lim_{V\to\infty}\frac{1}{V}\la b^*_q
b_q\ra_{H_{\sone,\sLambda}^{\rm eff}(\mu,\eta,\rho)}=0 , \ee i.e.,
there is \tit{no condensation} in the $q\neq 0$ modes and the laser
boson field.
\par
In this case the energy density is given by: \be
\lim_{V\to\infty}\frac{1}{V}\la
H_{\sone,\sLambda}(\mu)\ra_{H_{\sone,\sLambda}^{\rm
eff}(\mu,\eta,\rho)} =2\varepsilon_0(0)- \frac{\mu^2}{2\lambda} \ee
and the entropy density has the form: \be s(\mu)=2s_0(0)
=2\beta(\varepsilon_0(0)+p_0(0)). \ee Thus for the pressure one
gets: \be p(\mu)=2p_0(0)+\frac{\mu^2}{2\lambda}. \ee Notice that the
bound (\ref{eta-inequal}) and (\ref{eta-case-2}) imply the upper
limit on chemical potential
\begin{equation}\label{chem-poten-case-2}
\mu \leq \mu_c + \frac{4 \Omega\lambda \epsilon(q)}{g^2}
\end{equation}
for which Solution 2 applies.

This means that for the higher densities or chemical potentials:
\begin{equation}\label{mu-case-B}
\mu > \mu_c + \frac{4 \Omega\lambda \epsilon(q)}{g^2},
\end{equation}
to satisfy equation (\ref{density1}) we have to consider
\begin{equation}\label{Case-B}
Case\,\, B: \,\,\,\,\,\,\,\,\,\,\,\,\,\,\,\,\,\,\lim_{V\to\infty}
E_{-} =0 .
\end{equation}
Then (see (\ref{eta-inequal}) and Remark \ref{rem-2.2}) in the
thermodynamic limit: $|\eta|^2\to 4\Omega(\delta+\epsq)/g^2 $. In
fact, to obtain a finite limit in (\ref{consist1}) the corresponding
large-volume asymptotic should to be \be\label{eta-asymp}
|\eta|^2\approx \frac{4\Omega}{g^2}\(\delta+\epsq-\frac{1}{\beta
V\tau}\) \ee for some $\tau>0$. This implies that
\be\label{E-asympt-1} E_+ \to \Omega+\epsq-\delta,\ \ \ \ E_-
\approx \frac{\Omega}{\beta V \tau(\Omega+\epsq-\delta)} \ee and
(\ref{consist1}) becomes in the limit: \be \eta
\(1-\frac{g^2\tau}{4\delta\Omega}\)=\frac{gh}{2\delta} \,.
\label{eta eqtn} \ee The last equation gives \be
\tau=\frac{4\delta\Omega}{g^2}-\frac{2h\Omega}{g\eta}. \ee Taking
the limit $V\to\infty$ in (\ref{density1}) we get \be
\mu=\frac{4\lambda\Omega(\delta+\epsq)}{g^2}+\frac{4\lambda
\delta\Omega} {g^2}-\frac{2\lambda h\Omega}{g\eta}
+2\lambda\rho_0(-\delta)-\delta. \label{densityB} \ee We can also
check by using the diagonalization that in this case \be
\lim_{V\to\infty}\frac{1}{V}\la
a^*_{0-}a_{0-}\ra_{H_{\sone,\sLambda}^{\rm
eff}(\mu,\eta,\rho;h)}=|\eta|^2=\frac{4\Omega(\delta+\epsq)}{g^2},
\label{a0-a0-h} \ee \be \lim_{V\to\infty}\frac{1}{V}\la
a^*_{q+}a_{q+}\ra_{H_{\sone,\sLambda}^{\rm
eff}(\mu,\eta,\rho;h)}=\tau=\frac{4\delta\Omega}{g^2}-\frac{2h\Omega}{g\eta},
\label{aq-aq-h} \ee \be \lim_{V\to\infty}\frac{1}{V}\la b^*_q
b_q\ra_{H_{\sone,\sLambda}^{\rm
eff}}(\mu,\eta,\rho;h)=\frac{g^2|\eta|^2\tau}{4\Omega^2}=
(\delta+\epsq)\left(\frac{4\delta}{g^2}-\frac{2h}{g\eta}\right).
\label{bq-bq-h} \ee

We note here that if we take $\delta=0$ in (\ref{densityB}), then
for $h \rightarrow 0$ this expression gives the limiting value of
the chemical potential (\ref{chem-poten-case-2}) for  Solution 2.
\par
\textbf{Solution 3:}\ \ Let $\mu > \mu_c + 4 \Omega\lambda
\epsilon(q)/g^2$, see (\ref{mu-case-B}).  Now we take Case B and let
$h\to 0$. Then by (\ref{a0-a0-h}), (\ref{aq-aq-h}) and
(\ref{bq-bq-h}) we obtain a \tit{simultaneous} condensation of the
excited/non-excited bosons and the laser photons in the $q$-mode:
\be \lim_{V\to\infty}\frac{1}{V}\la
a^*_{0-}a_{0-}\ra_{H_{\sone,\sLambda}^{\rm
eff}(\mu,\eta,\rho)}=|\eta|^2=\frac{4\Omega(\delta+\epsq)}{g^2},
\label{a0-a0} \ee \be \lim_{V\to\infty}\frac{1}{V}\la
a^*_{q+}a_{q+}\ra_{H_{\sone,\sLambda}^{\rm
eff}(\mu,\eta,\rho)}=\tau=\frac{4\delta\Omega}{g^2}, \label{aq-aq}
\ee \be \lim_{V\to\infty}\frac{1}{V}\la b^*_q
b_q\ra_{H_{\sone,\sLambda}^{\rm
eff}(\mu,\eta,\rho)}=\frac{4(\delta+\epsq)\delta}{g^2}.
\label{bq-bq} \ee Equation (\ref{densityB}) becomes: \be
\mu=\frac{4\lambda\Omega(\delta+\epsq)}{g^2}+\frac{4\lambda
\delta\Omega}{g^2} +2\lambda\rho_0(-\delta)-\delta. \label{density3}
\ee Using the diagonalization of (\ref{eff-Ham-0}) one computes also
\be \lim_{V\to\infty}\frac{1}{V}\la
a^*_{q+}b_q\ra_{H_{\sone,\sLambda}^{\rm
eff}(\mu,\eta,\rho)}=\frac{4\delta\sqrt{\Omega(\delta+\epsq
)}}{g^2}. \label{aq-bq} \ee In this case the energy density is given
by: \bea \lim_{V\to\infty}\frac{1}{V}\la H_{\sone,\sLambda}
(\mu)\ra_{H_{\sone,\sLambda}^{\rm eff}(\mu,\eta,\rho)}
&=&(\epsq-\mu)\frac{4\delta\Omega}{g^2}-\mu\frac{4\Omega(\delta+\epsq)}{g^2}
- \frac{8\Omega \delta(\delta+\epsq )}{g^2}
\non\\
&&+\Omega \frac{4(\delta+\epsq)\delta}{g^2}+2\varepsilon_0(-\delta)
-2(\delta+\mu) \rho_0(-\delta)+\half \frac{(\delta+\mu)^2} {\lambda}\non\\
&=&\frac{4\Omega (\delta+\epsq
)\delta}{g^2}+2\varepsilon_0(-\delta)-\half \frac{(\delta+\mu)^2}
{\lambda}. \eea The entropy density is again given by \be
s(\mu)=2s_0(\mu-\lambda \rho) \ee and the pressure becomes \be
p(\mu)=2p_0(-\delta)+\half \frac{(\delta+\mu)^2} {\lambda}
-\frac{4\Omega (\delta+\epsq)\delta}{g^2}. \ee Notice that only
$|\eta|$ is determined and not the phase (\ref{phase}) of $\eta$ ,
that is that we can only determine the state up to a gauge
transformation in the $\sigma=-$ fields.
\par
The fact that here we have condensation in a state with non-zero
momentum is extremely significant and is related to the
\textit{spontaneous breaking} of translation invariance. We shall
examine this important aspect in Section 4.
\par
The Solutions 1, 2 and 3 represent possible equilibrium states. For
a given value of the chemical potential $\mu$, two or even three of
these solutions may be possible, see Figure 1 below. To distinguish
between them we have to compare the corresponding pressures to
determine which is maximum. The analysis, which is given in the next
subsection, involves a detailed study of the pressure. We find that
the situation is as described below.
\par
Let $\kappa= {8\Omega\lambda}/{g^2} - 1$ and
$\alpha=\epsq(\kappa+1)/2$. From the condition for thermodynamic
stability we know that $\kappa>0$. In this notation \textbf{Solution
2} applies for $\mu_c \leq \mu \leq \mu_c + \alpha$. Let $\delta_0$
be the unique value of $\delta\in [0,\infty)$ such that
$2\lambda\rho'_0(-\delta)=\kappa$ and let
$\mu_0=2\lambda\rho_0(-\delta_0)+\kappa \delta_0$. Note that
$\mu_0<2\lambda\rho_c$.\\
The case when $\mu_0 +\alpha\geq 2\lambda \rho_c$ is easy. In this
situation \textbf{Solution 1} applies for $\mu\leq2\lambda \rho_c$
and there exists $\mu_1>\mu_0 +\alpha$ (see definition after
(\ref{def-mu-1})) such that \textbf{Solution 2} applies for
$2\lambda\rho_c<\mu<\mu_1$ and \textbf{Solution 3} for $\mu\geq
\mu_1$.

When $\mu_0 +\alpha<2\lambda \rho_c$ the situation is more subtle.
In Subsection 2.3 we shall show that there exists $\mu_1>\mu_0
+\alpha$ (\ref{def-mu-1}), such that \textbf{Solution 3} applies for
$\mu\geq \mu_1$. However we are not able to decide on which side of
$2\lambda \rho_c$, the point $\mu_1$ lies. If $\mu_1>2\lambda
\rho_c$ the situation is as in the previous subcase, while if $\mu_0
+\alpha<\mu_1<2\lambda \rho_c$ the \tit{intermediate phase} where
\textbf{Solution 2} obtains is \tit{eliminated}. This the situation
is similar to \cite{PVZ2}, where one has $\alpha=0$.

Note that for $\nu<3$, \textbf{Solution 1} applies when $\mu<\mu_1$
and \textbf{Solution 3} when $\mu\geq\mu_1$.
\par
\subsection{The Pressure for Model 1}

This subsection is devoted to a detailed study of the pressure for Model 1 as a
function of the chemical potential $\mu$.
\par
Recall that $\delta$ is the limiting value of $\lambda \rho-\mu$, $\kappa= {8\Omega\lambda}/{g^2} - 1$ and
$\alpha=\epsq(\kappa+1)/2$. From above we have the following classification:
\par
{\bf Solution 1:}\ \  Here $\mu \leq \mu_c$. The density equation
\be \mu=2\lambda\rho_0(-\delta)-\delta \label{delta1} \ee has a
unique solution in $\delta$, denoted by $\delta_1(\mu)$ (previously
denoted by $\delta(\mu)$). Let \be
p_1(\delta,\mu):=2p_0(-\delta)+\frac{(\delta+\mu)^2}{2\lambda}. \ee
Then \be p(\mu)=p_1(\delta_1(\mu),\mu). \ee {\bf Solution 2:}\ \
Here $\mu > \mu_c$,  $\delta=0$ and the pressure is given by \be
p(\mu)=p_2(\mu):=2p_0(0)+\frac{\mu^2}{2\lambda}. \ee {\bf Solution
3:}\ The equation (\ref{density3}) can be re-written as \be
\mu=2\lambda\rho_0(-\delta)+\kappa \delta+\alpha. \label{delta3} \ee
Recall that $\delta_0$ is the unique value of $\delta\in [0,\infty)$
such that $2\lambda\rho'_0(-\delta)=\kappa$,
$\mu_0=2\lambda\rho_0(-\delta_0)+\kappa \delta_0$ and that
$\mu_0<2\lambda\rho_c$.
\par
Then for $\mu<\mu_0+\alpha$, equation (\ref{delta3}) has \tit{no}
solutions. For $\mu_0+\alpha\leq \mu \leq 2\lambda \rho_c +\alpha$
this equation has \tit{two} solutions: ${\tilde \delta}_3(\mu)$ and
$\delta_3(\mu)$, where ${\tilde \delta}_3(\mu)<\delta_3(\mu)$ if $\mu\neq \mu_0
+\alpha$, and ${\tilde \delta}_3(\mu_0 +\alpha)=\delta_3(\mu_0 +\alpha)$.
Finally for $ \mu > 2\lambda \rho_c +\alpha$ it has a
\tit{unique} solution $\delta_3(\mu)$. Let \be
p_3(\delta,\mu):=2p_0(-\delta)+\frac{\{(\delta+\mu)^2-(\kappa+1)\delta^2 -2\alpha
\delta\}}{2\lambda}. \ee
Then
\be
\frac{dp_3({\tilde
\delta}_3(\mu),\mu)}{d\mu}=\frac{{\tilde \delta}_3(\mu)+\mu}{\lambda}<
\frac{\delta_3(\mu)+\mu}{\lambda}=\frac{dp_3(\delta_3(\mu),\mu)}{d\mu} \ee for
$\mu\neq \mu_0 +\alpha$. Since $p_3({\tilde \delta}_3(\mu_0
+\alpha),\mu_0 +\alpha)=p_3(\delta_3(\mu_0 +\alpha),\mu_0 +\alpha)$, \be
p_3({\tilde \delta}_3(\mu),\mu)<p_3(\delta_3(\mu),\mu) \ee for $\mu_0+\alpha<
\mu \leq 2\lambda \rho_c +\alpha$. Therefore \be
p(\mu)=p_3(\delta_3(\mu),\mu)
\ee
for all $\mu\geq \mu_0 +\alpha$.
\par
Note that ${\tilde \delta}_3(2\lambda \rho_c +\alpha)=0$ so that \be
p_3({\tilde \delta}_3(2\lambda \rho_c +\alpha),2\lambda \rho_c
+\alpha)=p_1(\delta_1(2\lambda \rho_c),2\lambda \rho_c)=p_2(2\lambda
\rho_c)=2p_0(0)+2\lambda \rho^2_c. \ee therefore \be p_3(\delta_3(\mu_0
+\alpha),\mu_0 +\alpha)=p_3({\tilde \delta}_3(\mu_0 +\alpha),\mu_0
+\alpha)<2p_0(0)+2\lambda \rho^2_c. \label{P3<P2} \ee Also for
large $\mu$, $p_3(\delta_3(\mu),\mu)\approx
(\mu^2/2\lambda)((\kappa+1)/\kappa)$ while $p_2(\mu)\approx
(\mu^2/2\lambda)$, so that \linebreak\hfill
$p_3(\delta_3(\mu),\mu)>p_2(\mu)$ eventually. We remark finally that
the slope of $p_3(\delta_3(\mu),\mu)$ is greater than that of
$p_2(\mu)$,
\be
\frac{dp_3(\delta_3(\mu),\mu)}{d\mu}=\frac{\delta_3(\mu)+\mu}{\lambda}>
\frac{\mu}{\lambda}=\frac{dp_2(\mu)}{d\mu},
\ee
so that the
corresponding curves intersect at most once.
\par
The case $\alpha=0$, i.e. $\epsilon (q=0) = 0$, has been examined in
\cite{PVZ2}.
\par
For the case $\alpha>0$ we have two \tit{subcases}: \nl The subcase
$\mu_0 +\alpha\geq 2\lambda \rho_c$ is easy. In this situation
Solution 1 applies for $\mu\leq2\lambda \rho_c$. From (\ref{P3<P2})
we see that
\begin{equation}\label{def-mu-1}
p_3(\delta_3(\mu_0 +\alpha),\mu_0 +\alpha)<2p_0(0)+2\lambda
\rho^2_c<p_2(\mu_0 +\alpha
\end{equation}
and therefore from the behaviour for large $\mu$ we can deduce that
there exists $\mu_1>\mu_0 +\alpha$ such that Solution 2 applies for
$2\lambda\rho_c<\mu<\mu_1$ and Solution 3 for $\mu\geq \mu_1$.
\par
\begin{figure}[hbt]
\begin{center}
\hskip -0.5cm
\includegraphics[width=16cm]{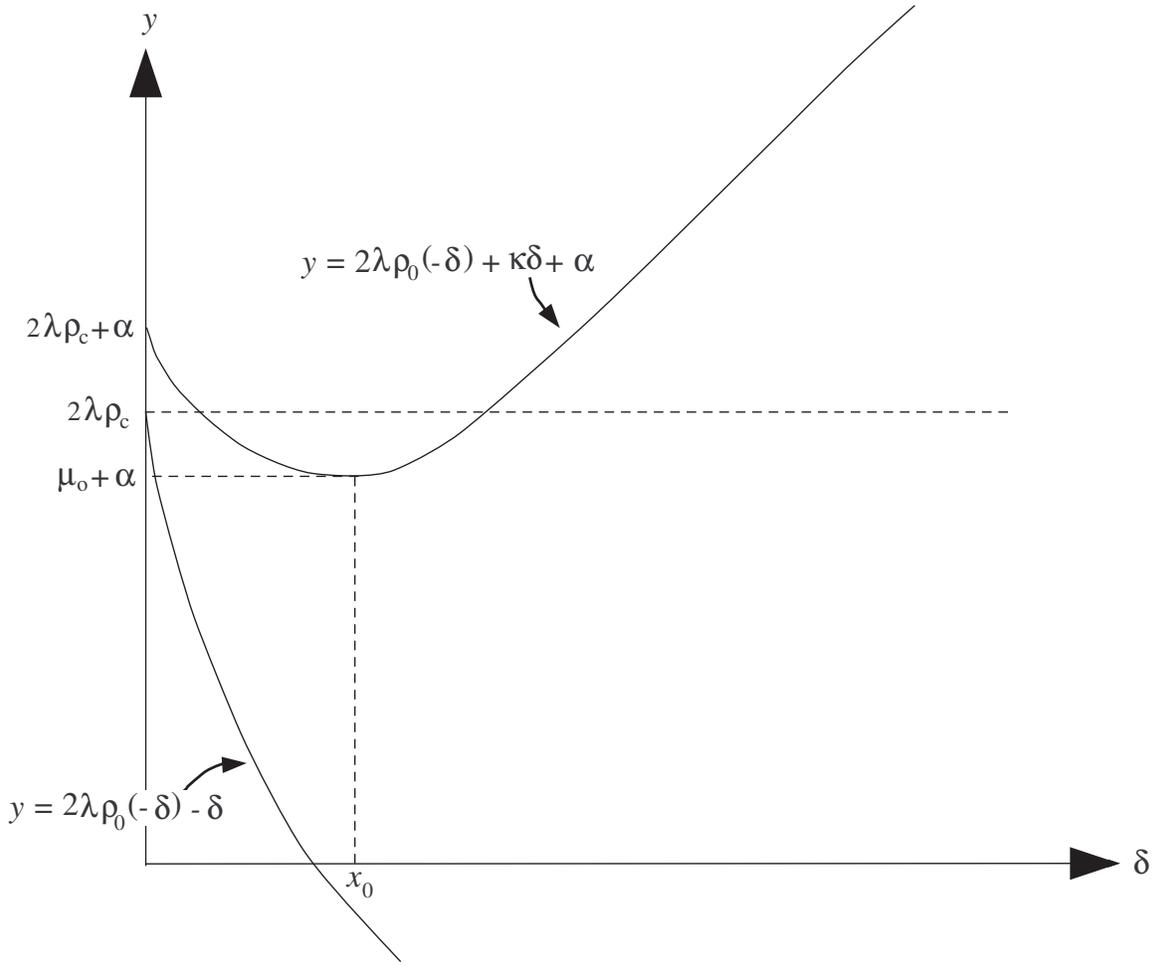}
\end{center}
\vskip- 1cm
\caption{The density equation for $h=0$ and $\nu \geq 3$} \label{a}
\end{figure}

The subcase $\mu_0 +\alpha<2\lambda \rho_c$ is more complicated. In
Figure \ref{a} we have drawn $y=2\lambda\rho_0(-\delta)-\delta$ and
$y=2\lambda\rho_0(-\delta)+\kappa \delta+\alpha$ for this subcase.
We know that
\be
p_3({\tilde \delta}_3(2\lambda \rho_c),2\lambda \rho_c)
<p_3({\tilde \delta}_3(2\lambda \rho_c +\alpha),2\lambda \rho_c
+\alpha)=p_1(\delta_1(2\lambda \rho_c),2\lambda \rho_c).
\ee
Therefore
since the slope of $p_3({\tilde \delta}_3(\mu),\mu)$ is greater than the
slope of $p_1(\delta_1(\mu),\mu)$ for $\mu_0 +\alpha<\mu<2\lambda
\rho_c$, (see Figure \ref{a}):
\be
 \frac{dp_3({\tilde
\delta}_3(\mu),\mu)}{d\mu}=\frac{{\tilde
\delta}_3(\mu)+\mu}{\lambda}>
\frac{\delta_1(\mu)+\mu}{\lambda}=\frac{dp_1(\delta_1(\mu),\mu)}{d\mu},
\ee we can conclude that \be p_3(\delta_3(\mu_0 +\alpha),\mu_0
+\alpha)=p_3({\tilde \delta}_3(\mu_0 +\alpha),\mu_0
+\alpha)<p_1(\delta_1(\mu_0 +\alpha),\mu_0 +\alpha). \ee We also
know by the arguments above that there exists $\mu_1>\mu_0 +\alpha$
such that Solution 3 applies for for $\mu\geq \mu_1$. However we do
know on which side of $2\lambda \rho_c$, the point $\mu_1$ lies. If
$\mu_1>2\lambda \rho_c$ the situation is as in the previous subcase
while if $\mu_0 +\alpha<\mu_1<2\lambda \rho_c$ the intermediate
phase where Solution 2 obtains is eliminated.
\section{Model 2}
As we said in the introduction the analysis for this model is very similar
to that of \tit{Model 1}. Therefore
we briefly summarize the results without repeating the details.
For \tit{Model 2} the effective Hamiltonian is
\bea
\label{eff-Ham-2} H_{\stwo,\sLambda}^{\rm eff}(\mu,\eta,\rho) &=&(\lambda \rho-\mu+\epsq)
a^*_q a_q+
(\lambda \rho-\mu)a^*_0a_0+\frac{g}{2}(\eta  a^*_q b_q +{\bar \eta }a_qb^*_q)\non\\
&&\hskip 1.5cm +\Omega\, b^*_qb_q+\frac{g\sqrt V}{2}\(\zeta
a_0+{\bar \zeta }a^*_0\)+T'_{\stwo,\sLambda} +(\lambda \rho-\mu)
N'_{\stwo,\sLambda} \eea where \be T'_{\stwo,\sLambda}=\sum_{k\in
\Lambda^*,\,k\neq 0\,k\neq q}\epsilon(k)N_k, \ee \be
N'_{\stwo,\sLambda}=\sum_{k\in \Lambda^*,\,k\neq 0\,k\neq
q}\epsilon(k)N_k. \ee The parameters $\eta$, $\zeta$ and $\rho$
satisfy the \tit{self-consistency} equations: \be \eta
=\frac{1}{\sqrt V}\la a_0 \ra_{H_{\stwo,\sLambda}^{\rm
eff}(\mu,\eta,\rho)},\ \ \ \ \zeta =\frac{1}{V}\la a^*_q
b_q\ra_{H_{\stwo,\sLambda}^{\rm eff}(\mu,\eta,\rho)} ,\ \ \ \ \rho
=\frac{1}{V}\la N_{\stwo,\sLambda}\ra_{H_{\stwo,\sLambda}^{\rm
eff}(\mu,\eta,\rho)}. \label{consist02} \ee Using the external
sources (\ref{h-sources}) and the same treatment as for the
\textit{Model 1} in Section 2.2, we again obtain three cases:
\par
{\bf Solution 1:}\ \ $\mu\leq \lambda\rho_c \equiv \mu_c$. In this
case there is \tit{no condensation}: \be
\lim_{V\to\infty}\frac{1}{V}\la a^*_0a_0\ra_{H_{\stwo,\sLambda}^{\rm
eff}(\mu,\eta,\rho)} =\lim_{V\to\infty}\frac{1}{V}\la a^*_q
a_q\ra_{H_{\stwo,\sLambda}^{\rm eff}(\mu,\eta,\rho)}
=\lim_{V\to\infty}\frac{1}{V}\la b^*_q
b_q\ra_{H_{\stwo,\sLambda}^{\rm eff}(\mu,\eta,\rho)} =0, \ee the
density equation is \be \mu=\lambda\rho_0(-\delta)-\delta
\label{density22} \ee and the pressure is \be
p(\mu)=p_0(-\delta)+\half \frac{(\delta+\mu)^2}{\lambda} . \ee
\par
\textbf{Solution 2:}\ \ Let $\mu_c<\mu \leq \mu_c + 4 \Omega\lambda
\epsilon(q)/g^2$. Then $\delta(\mu,h)=\lim_{h \rightarrow 0}(\lambda
\rho(\mu,h) - \mu)=0$. \be |\eta|^2= \lim_{V\to\infty}\frac{1}{V}\la
a^*_0 a_0\ra_{H_{\stwo,\sLambda}^{\rm eff}(\mu,\eta,\rho)}=
\frac{\mu}{\lambda}-\rho_c . \ee There is condensation in the $k=0$
mode but there is \tit{no condensation} in the $k=q$ mode and of the
photon laser field: \be \label{case-22}
\lim_{V\to\infty}\frac{1}{V}\la a^*_q
a_q\ra_{H_{\stwo,\sLambda}^{\rm eff}(\mu,\eta,\rho)} =
\lim_{V\to\infty}\frac{1}{V}\la b^*_qb_q\ra_{H_{\stwo,\sLambda}^{\rm
eff}(\mu,\eta,\rho)}=0 . \ee The pressure density is given by \be
p(\mu)=p_0(0)+\frac{\mu^2}{2\lambda}. \ee
\textbf{Solution 3:}\ \ Let $\mu > \mu_c + 4 \Omega\lambda
\epsilon(q)/g^2$, see (\ref{mu-case-B}). Then there is simultaneous
condensation of the zero-mode and the $q$-mode bosons as well as the
laser $q$-mode photons: \be \lim_{V\to\infty}\frac{1}{V}\la
a^*_0a_0\ra_{H_{\stwo,\sLambda}^{\rm eff}(\mu,\eta,\rho)}
=\frac{4\Omega(\delta+\epsq)}{g^2}, \label{a02-a02} \ee \be
\lim_{V\to\infty}\frac{1}{V}\la a^*_q
a_q\ra_{H_{\stwo,\sLambda}^{\rm eff}(\mu,\eta,\rho)}
=\frac{4\delta\Omega}{g^2}, \label{aq2-aq2} \ee \be
\lim_{V\to\infty}\frac{1}{V}\la b^*_q
b_q\ra_{H_{\stwo,\sLambda}^{\rm
eff}(\mu,\eta,\rho)}=\frac{4(\delta+\epsq)\delta}{g^2},
\label{bq2-bq2} \ee \be \lim_{V\to\infty}\frac{1}{V}\la a^*_q
b_q\ra_{H_{\stwo,\sLambda}^{\rm eff}(\mu,\eta,\rho)}
=\frac{4\delta\sqrt{\Omega(\delta+\epsq)}}{g^2}. \label{aq2-bq2} \ee
The density equation is \be
\mu=\frac{8\lambda\Omega}{g^2}(\delta+\epsq/2)+\rho_0(-\delta)-\delta
\label{density322} \ee and pressure is \be p(\mu)=p_0(-\delta)+\half
\frac{(\delta+\mu)^2}{\lambda} -\frac{4\Omega
(\delta+\epsq)\delta}{g^2}. \ee
\par
Note that relations between the values of $\mu$ in the three cases
above are exactly the same as for \tit{Model 1} apart from the
fact that $2\rho_0$ is now replaced by $\rho_0$ and $2\rho_c$ by
$\rho_c$. To see this one has to compare the kinetic energy operators
(\ref{kinet1}) and (\ref{kinet2}).
\section{Spontaneous Breaking of Translation Invariance and Matter-Wave Grating}
The recently observed phenomenon of \tit{ periodic spatial
variation in the boson-density} is responsible for the light and
matter-wave \tit{amplification} in superradiant condensation,
see \cite{I-K}-\cite{KI-1}, \cite{Koz}. This so called
\tit{matter-wave grating} is produced by the interference
of two different macroscopically occupied momentum states:
the first corresponds to a macroscopic number of \tit{recoiled}
bosons and the second to \tit{residual} BE condensate at rest.
Clearly this cannot happen in the translation invariant states and so it must be due
to a spontaneous breaking of this invariance.
\par
Let us consider the situation in Solution 3 for \tit{Model 1}. For
simplicity we shall take $q=(2\pi/\gamma) {\bf e}_1$, with ${\bf
e}_1=(1,0,\ldots,0)\in\RR^\nu $ and $\gamma>0$ and we shall denote
the limit Gibbs state for the effective Hamiltonian by $\omega$:
\be\label{Gibbs-state} \omega (\, \cdot\, ) =
\lim_{V\to\infty}\left\langle\, \cdot\, \right\rangle
_{H_{\sone,\sLambda}^{\rm eff}(\mu,\eta,\rho)}. \ee We know that the
existence of condensation in the zero-mode of the $\sigma=-$ bosons
implies that the extremal states have broken gauge symmetry in the
corresponding fields in this mode. As was remarked earlier this is
indicated here by the fact that $\eta$ is not zero.
\par
It has been shown in \cite{FPV} that condensation in the zero-mode implies that the gauge invariant
spatially homogeneous equilibrium states are not extremal but can be decomposed into a convex
combination of extremal space homogeneous equilibrium states with broken gauge symmetry
(spontaneous braking of gauge symmetry).
\par
From (\ref{aq-aq}), (\ref{bq-bq}) and (\ref{aq-bq}) we see that in
Solution 3 \be \lim_{V\to\infty}\left
|\frac{1}{V}\omega(a^*_{q+}b_q)\right |^2
=\lim_{V\to\infty}\frac{1}{V}
\omega(a^*_{q+}a_{q+})\lim_{V\to\infty}\frac{1}{V}\omega( b^*_q
b_q). \label{prod} \ee This strongly suggests a similar
decomposition when there is condensation in the $q$-mode. In fact
when there is condensation in a mode $q\neq0$ one can argue, along
the same lines as in \cite{FPV}, that the spatially homogeneous
equilibrium states when considered within the equilibrium states
which are periodic in the ${\bf e}_1$ direction with period
$\gamma$, are not extremal. They can be decomposed into a convex
combination of extremal periodic equilibrium states which are not
spatially homogeneous: \be\label{aver-per} \omega = \frac{1}{\gamma}
\int_0^\gamma dx\,\,\, \omega_x \ee where \be
\omega_x\circ\tau_{\gamma{\bf e}_1}=\omega_x \label{period1} \ee and
\be \omega_y\circ\tau_{x{\bf e}_1}=\omega_{(x+y)\hskip
-0.3cm\mod\gamma}. \label{period2} \ee Therefore in this model we
have spontaneous breaking of translation invariance. The rigorous
and explicit construction of the states $\omega_x$ involves
mathematical details which are outside of the scope of the present
paper and is carried out in \cite{PVZ3}.
\par
Let $\Lambda_1=\{x|x\in \Lambda, 0<x<\gamma\}$ and let
$V_1=|\Lambda_1|$. Then we can write, for example, \be
\lim_{V\to\infty}a^\# _{q+}/\sqrt V
=\lim_{N\to\infty}\frac{1}{2N}\sum_{n=-N}^N \frac{1}{V_1}\int_{n{\bf
e}_1+\Lambda_1} e^{\pm iqx}a^\#_+(x)dx. \ee Therefore in the
representation corresponding to each of the extremal states
$\omega_x$, $a^\# _{q+}/\sqrt V$ converges weakly to a complex
number which by (\ref{period1}) and (\ref{period2}) is equal to \be
e^{\mp iqx}\lim_{V\to\infty}\frac{1}{\sqrt V}\,\omega_0(a^\#_{q+}),
\ee where $\omega_0=\omega_{x=0}$. It then follows from
(\ref{aq-aq}), with $\delta=\lim_{V\to\infty}(\lambda \rho -\mu)$,
that \be \lim_{V\to\infty}\frac{1}{V}\omega_0(a^*_{q+}a_{q+}) =\left
|\lim_{V\to\infty}\frac{1}{\sqrt V}\,\omega_0(a_{q+})\right |^2
=\frac{4\delta\Omega}{g^2}. \label{aq-aq-per} \ee Similarly $a^\#
_{0-}/\sqrt V$ and $b^\# /\sqrt V$ converge weakly to a complex
numbers and from (\ref{a0-a0}) and (\ref{bq-bq}) respectively we
obtain \be \lim_{V\to\infty}\frac{1}{V}\omega_0(a^*_{0-}a_{0-})
=\left |\lim_{V\to\infty}\frac{1}{\sqrt V}\,\omega_0(a_{0-})\right
|^2 =\frac{4\Omega(\delta+\epsq)}{g^2} \label{a0-a0-per} \ee and \be
\lim_{V\to\infty}\frac{1}{V}\omega_0( b^*_q b_q) =\left
|\lim_{V\to\infty}\frac{1}{\sqrt V}\,\omega_0(b_q)\right |^2
=\frac{4(\delta+\epsq)\delta}{g^2}. \label{bq-bq-per} \ee The weak
convergence of
 $a^\# _{q+}/\sqrt V$ and $a^\# _{0-}/\sqrt V$ to complex numbers also implies
 that
\be\label{modes-limits1}
\lim_{V \to \infty}\frac{1}{V}\omega_0( a^*_{q,+}a_{k,+})=
\lim_{V \to \infty}\frac{1}{\sqrt{V}}\omega_0( a^*_{q,+})
\lim_{V \to \infty}\frac{1}{\sqrt{V}}\omega_0( a_{k,+})=0
\ee
for $k\neq q$ and
\be\label{modes-limits2}
\lim_{V \to \infty}\frac{1}{V}\omega_0( a^*_{0-}a_{k,-})=
\lim_{V \to \infty}\frac{1}{\sqrt{V}}\omega_0( a^*_{0,-})
\lim_{V \to \infty}\frac{1}{\sqrt{V}}\omega_0( a_{k,-})=0
\ee
for $k\neq 0$.
\par
An alternative strategy to the one we have developed above would be to use the traditional method of
introducing source terms in the Hamiltonian (\ref{Ham1}) to break both the gauge and
translation symmetries and let:
\be
\label{Ham1+sources}
H_{1,\Lambda}(\xi):= H_{1,\Lambda} - \sqrt{V}(\xi \,b^{*}_q +
\overline{\xi} \,b_q)  \,.
\ee
Then the effective Hamiltonian becomes
\bea
\label{eff-Ham1-sources}
H_{\sone,\sLambda}^{\rm eff}(\mu,\rho;\zeta,\xi) &=&(\lambda
\rho-\mu+\epsq) a^*_{q+}a_{q+}+ (\lambda \rho-\mu)a^*_{0-}a_{0-} \non\\
&& \hskip 0.7cm+\frac{g}{2}\left\{(\zeta + \frac{\xi}{\Omega})
a^*_{q+}a_{0-} + (\overline{\zeta}+
\overline{\frac{\xi}{\Omega}})a^*_{0-}a_{q+}\right\} + \Omega\,
\hat{b}^*_q \hat{b}_q \non \\
 && \hskip 1.5cm+V \Omega |\zeta|^2 - V
\frac{|\xi|^2}{\Omega} + T'_{\sone,\sLambda} +(\lambda \rho-\mu)
N'_{\sone,\sLambda} - \frac{1}{2}V \rho^2 \,,
\eea
where
\be
\label{hat-b-oper}
\hat{b}_q = b_q -\sqrt{V}(\zeta + \frac{\xi}{\Omega})
\ee
and we can carry out the procedure of Section 2.1 to obtain the results equivalent to
(\ref{aq-aq-per})-(\ref{modes-limits2}).
\par
We now want to examine the possibility of interference between two
different macroscopically occupied momentum states in \tit{Model 1}
in the periodic states.  Without loss of generality we can restrict
ourselves to the state $\omega_0$. In this state the mean local
particle density for the $\sigma=+$ bosons is \be
\label{part-dens-1} \rho_+(x)=
\lim_{V\to\infty}\frac{1}{V}\sum_{k\in\Lambda^*}
\sum_{p\in\Lambda^*} e^{i (k-p) x}\omega_0(a^*_{k,\, +}a_{p,\, +}) =
\rho +\lim_{V\to\infty}\frac{1}{V}\sum_{k\in\Lambda^*}
\sum_{p\in\Lambda^*\, p\neq k} e^{i (k-p) x}\omega_0(a^*_{k,\,
+}a_{p,\, +}). \ee We know that condensation occurs only in the
$q$-mode for the $\sigma=+$ bosons and in the $q=0$ - mode for the
$\sigma=-$ bosons and therefore only the terms containing $q$
survive in the integral sum (\ref{part-dens-1}) in the thermodynamic
limit: \be \label{part-dens2-1} \rho_+(x)= \rho
+\lim_{V\to\infty}\frac{1}{V}\sum_{k\in\Lambda^*\, k\neq q} \hskip
-0.3cm 2\,\Re{\rm \mathfrak{e}}\left\{e^{i (k-q) x}
\omega_0(a^*_{k,\, +}a_{q,\, +})\right\} \ee From the above
discussion, in particular from (\ref{modes-limits1}), we see that
$\rho_+(x)= \rho$. Similarly $\rho_-(x)= \rho$. Thus in spite of the
fact that the state $\omega_0$ is not space homogeneous, the total
particle density is constant and equal to $2\rho$. This means that
in \tit{Model 1} we get no particle density space variation even in
the presence of the \tit{light corrugated lattice of condensed
photons}, see (\ref{bq-bq-per}).
\par
Let us now turn our attention to the corresponding situation for
\tit{Model 2} in Solution 3. The decomposition into periodic states
still stands and again we have spontaneous breaking of gauge
symmetry. In the representation corresponding to each of the
extremal states, $a^\#_q /\sqrt V$, $a^\#_0 /\sqrt V$ and $b^\#
/\sqrt V$ all converge weakly to a complex numbers and \be
\lim_{V\to\infty}\frac{1}{V}\omega_0(a^*_q a_q) =\left
|\lim_{V\to\infty}\frac{1}{\sqrt V}\,\omega_0(a_q)\right |^2
=\frac{4\delta\Omega}{g^2}, \label{aq-aq-per2} \ee \be
\lim_{V\to\infty}\frac{1}{V}\omega_0(a^*_0a_0) =\left
|\lim_{V\to\infty}\frac{1}{\sqrt V}\,\omega_0(a_0)\right |^2
=|\eta|^2=\frac{4\Omega(\delta+\epsq)}{g^2}, \label{a0-a0-per2} \ee
\be \lim_{V\to\infty}\frac{1}{V}\omega_0( b^*_q b_q) =\left
|\lim_{V\to\infty}\frac{1}{\sqrt V}\,\omega_0(b_q)\right |^2
=\frac{4(\delta+\epsq)\delta}{g^2}. \label{bq-bq-per2} \ee We have
again \be \label{part-dens-2} \rho(x)=
\lim_{V\to\infty}\frac{1}{V}\sum_{k\in\Lambda^*}
\sum_{p\in\Lambda^*} e^{i (k-p) x}\omega_0(a^*_ka_p) = \rho
+\lim_{V\to\infty}\frac{1}{V}\sum_{k\in\Lambda^*}
\sum_{p\in\Lambda^*\, p\neq k} e^{i (k-p) x}\omega_0(a^*_ka_p). \ee
The important difference here is that in this model the \tit{same}
boson atoms may condense in two states and therefore \bea
\label{part-dens2-2} \rho(x)&=& \rho +\lim_{V\to\infty}\frac{1}{V}
2\,\Re{\rm \mathfrak{e}}\left\{e^{-iq x}
\omega_0(a^*_0a_q)\right\}\non\\
&& \hskip 1cm +\lim_{V\to\infty}\frac{1}{V}\sum_{k\in\Lambda^*\,
k\neq 0,q}\hskip -0.5cm 2\,\Re{\rm \mathfrak{e}}\left\{e^{i (k-q) x}
\omega_0(a^*_ka_q)\right\}\non\\
&&\hskip 2cm +\lim_{V\to\infty}\frac{1}{V}\sum_{k\in\Lambda^*\,
k\neq 0,q} \hskip -0.5cm 2\,\Re{\rm \mathfrak{e}}\left\{e^{i k x}
\omega_0(a^*_ka_0)\right\}. \eea The last two sums in
(\ref{part-dens2-2}) vanish in the thermodynamic limit by the same
argument as for \tit{Model 1}. However
\begin{equation}\label{0-q-corr}
\lim_{V \to \infty}\frac{1}{V}\omega_0( a^*_{q}a_{0}) =
\lim_{V \to \infty}\frac{1}{\sqrt{V}}\omega_0(a^*_{q})\lim_{V
\to \infty}\frac{1}{\sqrt{V}}\omega_0( a_{0}) \equiv C \neq
0 .
\end{equation}
Therefore, the bosons contained in the two condensates may interfere
and by virtue of (\ref{part-dens2-2}) and (\ref{0-q-corr}) this
gives the matter-wave grating formed by the quantum interference of
the two coherent states with different momenta: \be \label{grating}
\rho(x)=\rho+(C e^{i q x} + \overline{C} e^{- i q x}). \ee Notice
that by (\ref{0-q-corr}) and by (\ref{grating}) there is no
matter-wave grating in the Solution 2, when one of the condensates
(in this case the $q$-\tit{condensate}) is empty, see
(\ref{case-2}).
\section{Concluding Remarks}
We conclude this paper with few remarks concerning the
importance of the \tit{matter-wave grating} for the
amplification of light and matter waves observed in recent experiments.\\
It is clear that the absence of the matter-wave grating in \tit{Model 1}
and its presence in \tit{Model 2} provides a physical distinction between
Raman and Rayleigh superradiance. Note first that matter-wave
amplification differs from light amplification in one important
aspect: a matter-wave amplifier has to possess a \tit{reservoir}
of atoms. In \tit{Models 1} and \tit{2} this is the BE condensate. In both
models the superradiant scattering
transfers atoms from the condensate at rest to a recoil mode.\\
The \tit{gain} mechanism for the Raman amplifier is superradiant Raman
scattering in a two-level atoms, transferring bosons from the
condensate into the recoil state \cite{S-K}\\
The Rayleigh amplifier is in a sense even more effective. Since now
the atoms in a recoil state interfere with the BE condensate at
rest, the system exhibits a space \tit{matter-wave grating} and
the quantum-mechanical amplitude of transfer into the recoil state
is proportional to the product of the boson
occupation numbers $n_0 (n_q + 1)$ for the wave-vectors $k= 0,q$.
Each time the
momentum imparted by photon scattering is absorbed by the
matter-wave grating via the coherent transfer of an atom from the
condensate into the recoil mode. Thus, the variance of the grating
grows, since the quantum amplitude for scattered atom to be
transferred into a recoiled state is increasing
\cite{I-K}-\cite{KI-1}, \cite{Koz}. At the same time the dressing
laser beam prepares from the BE condensate a gain medium able to
amplify the light. The matter-wave grating diffracts the dressing
beam into the path of the probe light resulting in the amplification
of the latter \cite{KI-2}.
\par
In the case of equilibrium BEC superradiance the amplification of
the light and the matter waves manifests itself in \tit{Models 1}
and \tit{2} as a \tit{mutual enhancement} of the BEC and the photons
condensations, see Solutions 3 in Sections 2 and 3. Note that the
corresponding formul\ae\ for condensation densities for \tit{Model
1} (\ref{a0-a0})-(\ref{bq-bq}) and for \tit{Model 2}
(\ref{a02-a02}), (\ref{aq2-aq2}), (\ref{bq2-bq2}) are
\tit{identical}. The same is true for the \tit{boson-photon
correlations} (\tit{entanglements}) between recoiled bosons and
photons, see (\ref{aq-bq}), (\ref{aq2-bq2}), as well as between
photons and the BE condensate at rest:
\begin{equation}\label{a0-bq}
\lim_{V\to\infty}\frac{1}{V}\la a^*_{0-}
b_q\ra_{H_{\sone,\sLambda}^{\rm
eff}(\mu,\eta,\rho)}=\frac{1}{V}\la a^*_0
b_q\ra_{H_{\stwo,\sLambda}^{\rm eff}(\mu,\eta,\rho)} =\frac{4
\Omega(\lambda \rho+\epsq -\mu)\sqrt{\lambda \rho -\mu}}{g^2},
\end{equation}
and for the \tit{off-diagonal coherence} between recoiled atoms
and the condensate at rest:
\begin{equation}\label{a0-aq}
\lim_{V\to\infty}\frac{1}{V}\la a_{0-}a^*_{q+}
\ra_{H_{\sone,\sLambda}^{\rm eff}(\mu,\eta,\rho)}=\frac{1}{V}\la
a_0 a^*_{q}\ra_{H_{\stwo,\sLambda}^{\rm eff}(\mu,\eta,\rho)}
=\frac{4 \Omega\sqrt{(\lambda \rho+\epsq -\mu)(\lambda \rho
-\mu)}}{g^2} .
\end{equation}
As we have shown above, the \tit{difference} between
\tit{Models 1} and \tit{2} becomes visible only on the level
of the wave-grating or spatial modulation of the \tit{particle
density} (\ref{grating}).
\par
{\bf Acknowledgements:} Two of the authors (JVP and AFV) wish to
thank the Centre Physique Th\'eorique, CNRS-Luminy, where this work
was initiated, for its inspiring hospitality. They also wish to
thank Bruno Nachtergaele for his kind hospitality at the University
of California, Davis, where this work was continued. JVP wishes to
thank University College Dublin for the award of a President's
Research Fellowship.
%

\end{document}